\documentclass[twocolumn,showpacs,preprintnumbers,amsmath,amssymb,superscriptaddress]{revtex4}
\usepackage{epsfig,bm}
\usepackage{ulem}
\usepackage{amsmath}
\usepackage{color}

\begin{document}
\title{Harmonic-oscillator excitations of precise few-body wave functions}
\author{W. Horiuchi}
\affiliation{Department of Physics, Hokkaido University, Sapporo 060-0810, Japan}
\author{Y. Suzuki}
\affiliation{Department of Physics, Niigata University, Niigata 950-2181, Japan}
\affiliation{RIKEN Nishina Center, Wako 351-0198, Japan}
\pacs{
21.60.De, %Ab initio methods
21.30.-x, %Nuclear forces
27.10.+h %A<=5 
}

\begin{abstract}
A method for calculating the occupation probability
of the number of harmonic oscillator (HO) quanta is developed 
for a precise few-body wave function obtained in a correlated Gaussian basis. 
The probability distributions 
of two- to four-nucleon wave functions obtained using 
different nucleon-nucleon ($NN$) interactions are analyzed to 
gain insight into the characteristic behavior of the various interactions. 
Tensor correlations as well as short-range correlations
play a crucial role in enhancing
the probability of high HO excitations.
For the excited states of $^{4}$He, the 
interaction dependence is much less 
because high HO quanta are mainly responsible for  
describing the relative motion function 
between the $3N+N$ ($^3$H$+p$ and $^3$He$+n$) clusters.
\end{abstract}
\maketitle

\section{Introduction}
The nuclear shell model is a standard microscopic theory for describing 
a many-nucleon system. For doubly closed nuclei, 
first we consider the lowest HO
state expressed with a single Slater determinant (SD), 
an antisymmetrized product of single-particle HO orbits.
To take many-body correlations into account,
multi-particle-multi-hole ($m$p-$m$h) configuration mixing calculations are  
performed by superposing many SD states that include higher HO excitations.

Thus far, the {\it ab initio} no-core shell model (NCSM) with
modern nuclear forces has been developed extensively~\cite{barrett13}.
In the NCSM, all nucleons are active, 
but a crucial approximation is the truncation of maximum HO quanta, 
which determines the NCSM space.
Compared to ordinary shell-model effective interactions, the 
use of realistic nuclear forces requires
many SD states in higher major shells to reach convergence 
because of strong couplings between
low- and high-momentum components arising from the 
tensor component and short-range repulsion of the nuclear force.

The HO expansion provides us with 
systematic and  size extensive calculations,
but it is not advantageous to describe 
the spatial extent of the wave function
because of a rapid fall-off of the HO wave functions.
If a system exhibits a cluster structure, the 
subsystems are localized and their relative motion wave function
can have a long tail.
For example, the first excited state of $^{16}$O
is still difficult to reproduce by present large-scale 
shell-model calculations~\cite{wloch05,maris09}.
As shown in Refs.~\cite{suzuki76,suzuki96,neff09,horiuchi14},
the cluster structure always induces extremely high HO quanta
beyond the limitation of present computational resources.
To make an efficient description of nuclear many-body systems 
using the HO expansion, it is useful to know how many  
HO excitations are required to account for 
those important correlations which are induced
by the tensor force, the short-range repulsion, and the cluster structure.

Nowadays, precise wave functions of very light nuclei 
can be obtained using {\it ab initio} few-body methods.
Though the few-body method can only be applied to very light nuclei,
long-ranged asymptotics as well as 
short-range correlations are accurately described.
Such few-body wave functions can provide
important hints on how to tackle 
complicated many-body problems.
Since the shell model and the few-body model 
are formulated in different frameworks,
it is convenient to relate the few-body wave function to 
the HO wave function. For this purpose, 
by extending the formulation of Ref.~\cite{suzuki96}, we 
calculate the occupation probability of the number of total HO quanta 
in the wave function obtained with an {\it ab initio} few-body method,  
in particular, using a correlated Gaussian (CG) basis 
with global vectors~\cite{varga95,svm,suzuki08,aoyama12}.
We calculate the HO occupation probability of the wave functions of
two- to four-nucleon systems and discuss its properties, 
especially focusing on the $^4$He system.

The nucleus $^4$He is an interesting example
because important ingredients of many-nucleon dynamics show up in its spectrum.
The ground state is strongly correlated due to the tensor
component and short-range repulsion of the nuclear force.
The tensor force induces 
a $D-$state mixing of approximately 14\%~\cite{kamada01}.
Since the two nucleons cannot overlap with each other
due to the short-range repulsion, 
the universal short-ranged behavior is found 
in the pair correlation functions~\cite{forest96,feldmeier11}.
In the excited states, the structure changes drastically.
The first excited $0_2^+$ state of $^4$He is recognized to have a
$^3$H$+p$ and $^3$He$+n$ ($3N+N$) cluster structure
with $S$-wave relative motion~\cite{hiyama02}.
The negative-parity states are located at just a few MeV 
above the $0_2^+$ state
and are considered to be parity-inverted partners 
of the $0_2^+$ state. They have the intrinsic structure 
of $3N+N$ with $P$-wave relative motion~\cite{horiuchi08}.
Thus, $^4$He has a similarity to $^{16}$O~\cite{suzuki76} that 
exhibits a $^{12}{\rm C}+\alpha$ cluster structure in its spectrum.

The analysis of the oscillator excitations will be useful for 
developing and improving a truncation scheme for the model space in,  
e.g., {\it ab initio} NCSM~\cite{barrett13},
importance-truncated NCSM~\cite{roth07,forssen13},
symmetry-adapted no-core shell model~\cite{dytrych07}, 
Monte Carlo shell model~\cite{shimizu13} as well as a model approach like 
the tensor-optimized shell model~\cite{myo07}.

The paper is organized as follows. Section~\ref{count.sec} 
gives a basic formula to calculate the occupation probability of the HO quanta in 
the CG basis. 
Section~\ref{fewbody.sec} defines the Hamiltonian and the CG basis, 
and briefly explains how we obtain the precise few-body wave functions.
Section~\ref{results.sec} demonstrates the HO occupation 
probability distributions for two- to four-nucleon systems.
Four different potentials are employed in order to examine
how the HO distribution reflects the characteristics of the nuclear force.
In Sec.~\ref{ground.sec}, we discuss the role of the HO quanta
higher than the lowest $(0s)^N$ configuration,
focusing on the relationship with
the short-range repulsion and tensor correlations in the ground state of $^4$He.
Long-ranged cluster correlations in the excited states of $^4$He
are discussed in Sec.~\ref{excited.sec}. Section~\ref{inversion.sec} 
discusses the parity-inverted partners of the first excited state of $^4$He.
The summary is given in Sec.~\ref{summary.sec}. 
An appendix details a method for calculating the probability of the HO quanta. 

\section{Formulation}
\label{formulation.sec}

\subsection{Harmonic-oscillator occupation probability}
\label{count.sec}

Let $\Psi$ denote an $N$-nucleon wave function that is spurious center-of-mass 
(c.m.) motion free. 
The occupation probability $P_Q$ of the $Q \, \hbar \omega$ components in $\Psi$ is calculated using  
an integral of the projection-operator type   
\begin{align}
P_Q=\frac{1}{2\pi}\int_{0}^{2\pi} d\theta \, \text{e}^{-iQ\theta}
\left<\Psi\right|\text{e}^{i\theta\mathcal{O}}\left|\Psi\right>,
\label{HOquanta.eq}
\end{align}
where $\mathcal{O}$ is an operator that counts the number of HO quanta 
\begin{align}
\mathcal{O}=\sum_{i=1}^{N-1}\left(\frac{1}{\hbar\omega}H_{\rm HO}(i)-\frac{3}{2}\right).
\label{HOope.eq}
\end{align}
Here $H_{\rm HO}(i)$ is the HO Hamiltonian for the $i$th relative coordinate. See the appendix for details. 
The method for calculating $P_{Q}$ was developed 
for microscopic multicluster wave functions in Ref.~\cite{suzuki96}.
We extend it to the case where $\Psi$ is given in terms of a combination of 
CG basis functions.

The CG is constructed from the 
generating function~\cite{varga95,svm} 
\begin{align}
g(\bm{s}; A,\bm{x})=\exp\left(-\frac{1}{2}\tilde{\bm{x}}A\bm{x}+\tilde{\bm{s}}\bm{x}\right),
\label{gCG.eq}
\end{align}
where the $(N-1)\times(N-1)$ matrix $A$ is  positive-definite and symmetric, 
$\bm{s}$ is a  column vector of an $(N-1)$ dimension to describe the angular motion of the system, and 
$\bm{x}$ is a column vector of an $(N-1)$ dimension  whose element
is the 3-dimensional relative coordinate $\bm x_i$. The transpose of a matrix is indicated by a tilde symbol. 
Both $A$ and $\bm s$ are variational parameters, which makes 
 the CG flexible and easily adapted to few-body problems. 
To calculate the quantity (\ref{HOquanta.eq}) with the CG,
we start from the matrix element of $\text{e}^{i\theta\mathcal{O}}$
between the generating functions~(\ref{gCG.eq}).  As detailed in 
the appendix, the required matrix element reads 
\begin{align}
&\left<g(\bm{s}^\prime;A^\prime,\bm{x})\right|\text{e}^{i\theta\mathcal{O}}\left|
g(\bm{s};A,\bm{x})\right>\notag\\
&=\left(\frac{(2\pi)^{N-1}\det\Gamma}{\det B\det C}\right)^{\frac{3}{2}}
\exp\left(-\frac{1}{2}\tilde{\bm{s}}G\bm{s}+\frac{1}{2}\tilde{\bm{v}}B^{-1}\bm{v}\right),
\label{formulaQ.eq}
\end{align}
where $B=D+A^\prime$, $\bm{v}=z\Gamma C^{-1}\bm{s}+\bm{s}^\prime$,
and $z=\text{e}^{i\theta}$. The matrices $\Gamma, C, D$, and $G$ are defined in the appendix.  
Once the matrix element between the CG bases is obtained as a function of $\theta$,
the integration in Eq. (\ref{HOquanta.eq}) is performed numerically.

\subsection{Few-body wave functions}
\label{fewbody.sec}

\subsubsection{Hamiltonian}

The Hamiltonian of the $N$-nucleon system 
is composed of the kinetic energy, two-body $NN$ 
interaction, and three-body interaction (3NF) terms 
\begin{align}
H=\sum_{i=1}^{N}T_i-T_{\rm cm}+\sum_{i<j}v_{ij}+\sum_{i<j<k}v_{ijk}.
\end{align}
The c.m. kinetic energy is subtracted and no spurious c.m. motion 
is involved in the calculation.
The inputs used in this paper are
$\hbar^2/m=41.47106$ MeV\,fm$^2$ and $e^2=1.440$ MeV\,fm. 
The proton mass and neutron mass are taken to be equal.

We adopt (i) Minnesota (MN)~\cite{MN}, (ii) Afnan-Tang S3 (ATS3)~\cite{ATS3}, 
(iii) G3RS~\cite{G3RS}, and 
(iv) AV8$^\prime$~\cite{AV8p} potential models as the $NN$ interaction. 
A central 3NF~\cite{hiyama02} is added together with 
the realistic G3RS and AV8$^\prime$ potentials in order 
to reproduce the binding energies of $^3$H and $^4$He.  
The MN potential is often used in  
microscopic cluster-model calculations.
Though it has only a central term, the potential reproduces 
the binding energies of $N=2-6$ systems fairly well~\cite{varga95}.
The ATS3 potential also has only the central term 
but contains a strong short-range repulsive core.
The AV8$^\prime$ potential consists of central, spin-orbit, and tensor
components, as well as has  strong short-range repulsion.
The G3RS potential is somewhat softer than AV8$^\prime$
and gives a smaller $D$-state probability.
The $\bm{L}^2$ and quadratic $\bm{L}\cdot\bm{S}$ terms in the G3RS
potential are ignored. 

\subsubsection{Correlated Gaussians and global vectors}

The wave function $\Psi$ is given as a combination of 
the basis functions expressed in the $LS$ coupling scheme
\begin{align}
\Phi_{(LS)JM_JTM_T}=\mathcal{A}\left[F_{L}(\bm{x})\chi_S^{\text{(spin)}}\right]_{JMJ}\eta_{TM_T}^{\text{(isospin)}},
\label{basisfn}
\end{align}
where $\mathcal{A}$ is the antisymmetrizer,
and the square brackets, $[\dots]$, denote the angular momentum coupling.
The spin function is given in a successive coupling scheme 
\begin{align}
\chi^{\text{(spin)}}_{SM_S}=
[\dots[[\chi_{1/2}(1)\chi_{1/2}(2)]_{S_{12}}
\chi(3)]_{S_{123}}\dots]_{SM_S}.
\end{align}
The isospin wave function has exactly the same form as the spin part.
All possible intermediate spins and isospins are included in the basis set.
The orbital part is represented by
the CG with  
two global vectors
\begin{align}
&F_{(L_1L_2)LM_L}(u_1,u_2,A,\bm{x})\notag\\
&=\exp\left(-\frac{1}{2}\tilde{\bm{x}}A\bm{x}\right)
\left[\mathcal{Y}_{L_1}(\tilde{u}_1\bm{x})\mathcal{Y}_{L_2}
(\tilde{u}_2\bm{x})\right]_{LM_L}
\label{GVR.eq}
\end{align}
with a solid harmonic
\begin{align}
\mathcal{Y}_{\ell m}(\bm{r})=r^\ell Y_{\ell m}(\hat{\bm{r}}),
\end{align}
where $u_i$ is an $(N-1)$-dimensional column vector
and $\tilde{u}_i\bm{x}$ is called a  global vector that 
describes the rotational motion of the system.
The off-diagonal matrix elements of $A$  explicitly 
describe correlations among the particles.
The matrix element of the Hamiltonian 
between the CG of Eq. (\ref{GVR.eq}) can be obtained analytically
from the one between the generating functions (\ref{gCG.eq})
in a systematic, algebraic procedure prescribed in Refs.~\cite{svm,suzuki08,aoyama12}.
The CG basis~(\ref{GVR.eq}) has the great advantage that its functional form
remains unchanged under a coordinate transformation, thereby facilitating easily 
operations such as the ones involved in channel rearrangements and permutations, etc.
This flexibility enables us to apply the CG approach to many quantum-mechanical 
few-body problems.
See Ref.~\cite{mitroy13} for recent various applications of the CG.

The ground states of $^{2,3}$H, $^4$He and the excited states of $^4$He
are obtained using the stochastic variational method~\cite{varga95,svm}.
Though all the excited states of $^{4}$He are above the
$^3$H+$p$ threshold,  
we describe them in the square-integrable CG basis functions. 
Since they have relatively small decay
widths ranging from 0.5 to 2\,MeV~\cite{tilley92},
the bound-state approximation works reasonably well as discussed
in Ref.~\cite{horiuchi13a}.
More details of calculations are given in 
Refs.~\cite{suzuki08,horiuchi08,horiuchi13b}.

\section{Results and discussions}
\label{results.sec}

\subsection{Ground states: tensor and short-range correlations}
\label{ground.sec}

\begin{figure}[ht]
\begin{center}
\epsfig{file=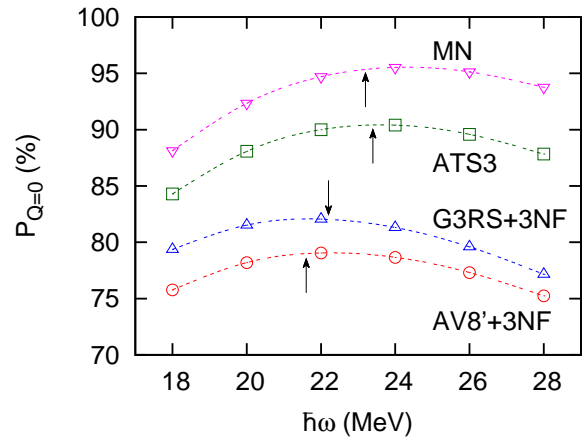,scale=1.4}
\caption{(Color online) $\hbar\omega$ dependence of the occupation probability
of the lowest HO quantum in the ground state of $^4$He.
The arrow indicates the $\hbar\omega$ value that is employed to calculate 
$P_Q$ values of $^4$He. See text for detail.}  
\label{HOhw4He.fig}
\end{center}
\end{figure}

Figure~\ref{HOhw4He.fig} displays
the probability of the lowest HO quantum, $P_{0}$, 
for the ground state of $^4$He as a function of the oscillator frequency $\hbar\omega$.
We see moderate $\hbar\omega$ dependence of $P_{0}$ 
in all the potential models. Since the $P_Q$ distribution depends on $\hbar \omega$, 
we fix it by requiring that the lowest shell-model 
configuration, $(0s)^N$, for the fixed $\hbar \omega$ reproduces the root-mean-square (rms) matter radius of the precise wave function. This is reasonable because the $(0s)^N$ configuration 
is the dominant component of the wave function for
$s$-shell nuclei. 
The $\hbar \omega$ values determined for $^4$He are 23.2, 23.4, 22.2, and 21.6\,MeV for 
MN, ATS3, G3RS+3NF, and AV8$^\prime$+3NF potentials, respectively.
The $P_{0}$ values calculated with these $\hbar \omega$ values are 
close to the maximum values in Fig.~\ref{HOhw4He.fig}.

\begin{table}[ht]
\begin{center}
\caption{Ground-state energies ($E$) and rms matter radii ($r_m$) of
two- to four-nucleon systems calculated with different potential models:
(i) MN, (ii) ATS3, (iii) G3RS+3NF, and (iv) AV8$^\prime$+3NF potentials.  
The $D$-state probability $P(D)$ and the occupation probability $P_{0}$ of the
$0\,\hbar \omega$ component are given in percents.
$M_Q$ and $\sigma_Q$ denote the average and standard
deviation of the $P_Q$ distribution. 
}
\begin{tabular}{ccccccccccccc}
\hline\hline  
&     &&$E$(MeV) &$r_m$(fm)&$P(D)$(\%)&$P_{0}$(\%)&$M_Q$&$\sigma_Q$\\
\hline
$^2$H       &(i)         &&$-$2.20&1.95&0.00&89.6&0.534&1.95\\
($1^+0$) &(ii)          &&$-$2.22 &1.94&0.00&89.4&0.692&3.77\\
            &(iii)      &&$-$2.28&1.98&4.78&86.9&1.27&5.80\\
            &(iv)&&     $-$2.24&1.96&5.77&85.5&1.57&6.84\\
\hline
$^3$H &(i)            &&$-$8.38 &1.71&0.00&90.8&0.409&1.70\\
($\tfrac{1}{2}^+\tfrac{1}{2}$)
      &(ii)          &&$-$8.76 &1.67&0.00&89.7&0.787&3.99\\
      &(iii)      && $-$8.35&1.74  &7.10&84.9&1.52&5.96\\
      &(iv)&&   $-$8.41  &1.70&8.69&83.1&1.92&7.08\\
\hline
$^4$He  &(i)            &&$-$29.94&1.41&0.00&95.4& 0.263&1.48\\
($0_1^+0$)&(ii)          &&$-$30.83&1.42&0.00& 90.4&0.934&4.00        \\
      &(iii)      &&$-$28.56     &1.47 &11.42& 82.1&1.96&5.98        \\
      &(iv)&&$-$28.43       &1.45 &14.07&79.1& 2.59&7.31       \\
\hline\hline
\end{tabular}
\label{quanta.tab}
\end{center}
\end{table}

Table~\ref{quanta.tab} summarizes the calculated energy $E$, rms 
matter radius $r_m$, $D$-state probability $P(D)$, 
and $P_{0}$ of the ground state of $^2$H, $^3$H, and $^4$He  
for the different potential models.
All the interactions give approximately  the same $E$ and $r_m$ 
but quite different $P(D)$. The MN potential, which is the softest among the four potentials, 
gives the largest $P_{0}$ of approximately 95\% for $^4$He.
The $N=2-4$ wave functions with the MN potential 
are well described by the $(0s)^N$ configurations.
When the other interactions are employed, the mixing of higher-$Q$ 
components becomes important.
When a realistic potential is used,
the deviation from the $(0s)^N$ structure is the largest in $^{4}$He, which   
is the most tightly bound and 
has the largest $D$-state probability, as a result of
the effects of short-range and tensor correlations.
The ground state of $^4$He obtained with the AV8$^\prime$+3NF 
interaction predicts at most 80\% of the $(0s)^4$ configuration.

\begin{figure}[ht]
\begin{center}
\epsfig{file=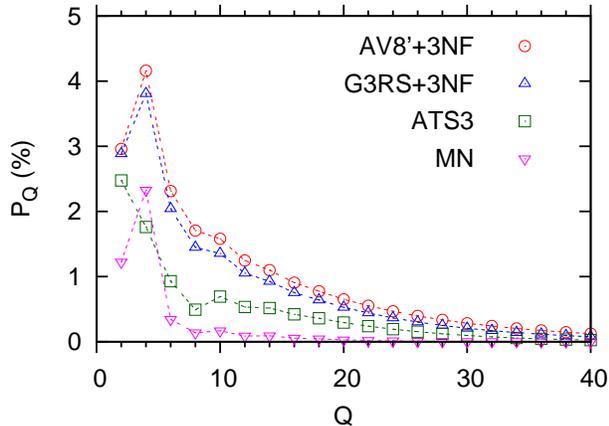,scale=1.35}
\caption{(Color online) Occupation probability distributions 
of the total HO quanta for
the ground state of $^4$He calculated with different potential models.
$P_{0}$ is not shown in the figure but given in Table~\ref{quanta.tab}.  
$P_Q$ values are connected by dotted lines to guide the eye.
}
\label{HOquanta4He.fig}
\end{center}
\end{figure}

Figure~\ref{HOquanta4He.fig} plots $P_{Q}\, (Q>0)$ of $^4$He. 
Consistently with the $M_Q$ and $\sigma_{Q}$ values in  Table~\ref{quanta.tab}, 
a harder interaction leads to $P_{Q\neq 0}$ that is more 
enhanced and extended to larger $Q$. 
In the case of the MN potential,
$P_Q$  is found to be about 1\% to 2\% for
$Q=2$ and $4$, but it diminishes rapidly with increasing $Q$.
Since no short-range repulsion is present in the MN potential,
the configurations contributing to $P_{2}$ and $P_{4}$, e.g.,   
$(0s)^3(1s)$ for $Q=2$ and $(0s)^2(1s)^2$, $(0s)^3(2s)$ for $Q=4$  
are expected to improve the tail of 
the wave function that cannot be described with $(0s)^4$ alone.
One may wonder why $P_{2}$ is smaller than $P_{4}$.
We recalculate $P_Q$ using smaller $\hbar\omega$
to describe the tail part more efficiently.
For $\hbar\omega$ less than 20 MeV, the $P_Q$ distribution shows a 
monotonous decrease with increasing $Q$. 
The $P_Q$ values for small $Q$ depend on the choice of $\hbar\omega$.
We will discuss this later in this section. With the ATS3 potential, 
$P_Q$ decreases monotonously up to $Q=8$ 
and exhibits a bump at $Q=10$ with a long tail 
extending to more than $Q=30$,  which is apparently  due to the short-range repulsion.
The G3RS+3NF and AV8$^\prime$+3NF potentials give a very similar pattern characterized by 
large and very extended $P_Q$ distributions.
The probability is still 1.6\% at $Q=10$ and
0.7\% at $Q=20$ when the AV8$^\prime$+3NF potential is used.

\begin{figure}[ht]
\begin{center}
\epsfig{file=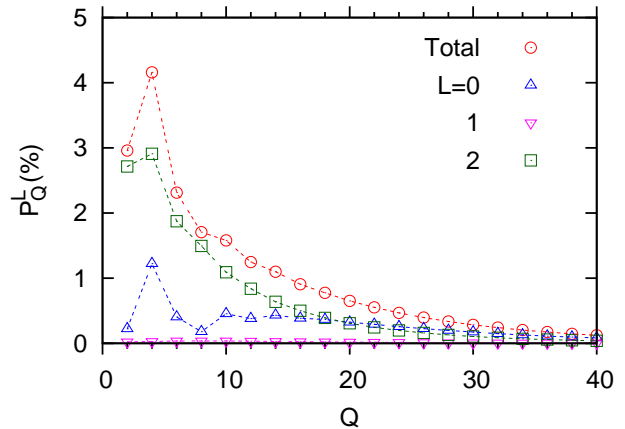,scale=1.35}
\caption{(Color online) Angular momentum decomposition 
of the occupation probability of the total HO quanta
for the ground state of $^4$He. $P^L_0$ values in percents are 79.1, 0, 0 for $L=0, 1, 2$, respectively. The AV8$^\prime$+3NF potential is used.}
\label{HOquantaAV8pdec.fig}
\end{center}
\end{figure}

To discuss whether the short-range repulsion or the tensor component in the 
$NN$ interaction is important in determining the $P_Q$ distribution, 
we decompose $P_Q$ according to the total orbital angular momentum $L$.
The ground-state wave function of $^4$He is expressed in the 
notation of Eq.~(\ref{basisfn}) as
\begin{align}
\Psi=\sum_{L=0,1,2}\sum_{i}C_L^{(i)}\Phi_{(LL)0000}^{(i)},
\end{align}
where the amplitude $C_L^{(i)}$ of the $i$th basis state $\Phi_{(LL)0000}^{(i)}$  
satisfies $\sum_{L=0,1,2}\sum_i (C_L^{(i)})^2=1$. The $P_Q$ is decomposed to a sum of 
$P_Q^L$ that is defined by  
\begin{align}
P_Q^L&=\frac{1}{2\pi}
\sum_{i}(C_L^{(i)})^2\int_{0}^{2\pi} d\theta \, \text{e}^{-iQ\theta}\notag\\
&\times
\left<\Phi_{(LL)0000}^{(i)}\right|\text{e}^{i\theta\mathcal{O}}\left|\Phi_{(LL)0000}^{(i)}\right>.
\end{align}

Figure~\ref{HOquantaAV8pdec.fig} displays $P_Q^L$ of 
the ground state of $^4$He calculated with the AV8$^\prime$+3NF potential.
The $P^1_Q$ is negligibly small
because the $L=1$ component occupies only 0.37\% 
of the total wave function~\cite{horiuchi13b}.
The $L=2$ component can couple with the $L=0$ configurations 
through the tensor force that induces a major shell mixing in the wave function.
The $P_Q^{2}$ dominates up to $Q=18$, where the
$P_Q^{0}$ gives an equal contribution.
The $P^{0}_Q$ distribution shows a bump at $Q=10$ with a long tail 
similarly to the ATS3 potential case.
This  suggests that the bump and tail behavior 
in the $L=0$ component is due to the short-range repulsion.
Both the tensor and short-range characters of the $NN$ potential
make the convergence 
of conventional shell-model calculations very slow.

\begin{figure}[ht]
\begin{center}
\epsfig{file=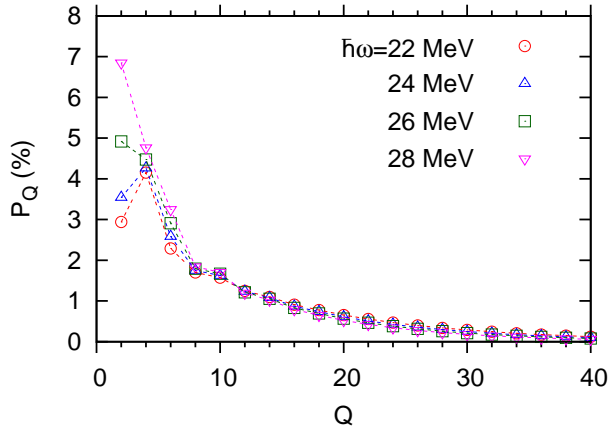,scale=1.35}
\caption{(Color online) $\hbar\omega$ dependence
of the occupation probability of the total HO quanta with $Q>0$
for the ground state of $^4$He. The AV8$^\prime$+3NF potential is used.}
\label{HOhwdepAV8p.fig}
\end{center}
\end{figure}

Figure~\ref{HOhwdepAV8p.fig} presents how the probability distribution changes 
with different $\hbar\omega$ values. 
Though $P_{2}$ depends on $\hbar\omega$, the dependence of 
the sum of $P_{0}$ and $P_{2}$ is much weaker. This is understood as follows. 
Since the main role of the configurations with $Q=0$ and 
2 is considered to describe the mean-field correlation of the system, each of 
$P_Q$ values may depend on a choice of $\hbar \omega$ 
but the sum of them may not so much. 
A weaker dependence of $P_Q$ at $Q=4$ and 6 reflects 
the dominance of the tensor correlations. 
Finally no $\hbar\omega$ dependence is found for $Q>6$.
The higher-$Q$ components are always
present and remain unchanged for different choices of $\hbar\omega$.

The mechanism responsible for enhancing the high-$Q$ components is different 
for the short-range repulsion and the tensor correlations
of the realistic $NN$ interaction. 
The total number of HO quanta $Q$ is a sum of the HO quanta,   
$\sum_{i=1}^N (2n_i+l_i)$, where $n_i$ and $l_i$ are
respectively the principal and azimuthal quantum numbers 
of the HO wave function of the 
$i$th nucleon. Since no spurious c.m. motion is included, 
the sum ranges over all the nucleons.  
As shown in Refs.~\cite{forest96,kamada01,feldmeier11},
the short-range repulsion makes
a strong depression at short distances
in the pair correlation function.
In the HO expansion, this depression of the pairwise relative 
wave functions at short distances  
is taken care of 
by superposing many HO wave functions that 
have larger $n_i$ with the same $l_i$, which obviously leads 
to the large-$Q$ components. On the other hand, 
the tensor correlations induce  high-$Q$  components,
because of the couplings between the HO wave functions with different $l_i$.

The $P_Q$ distribution actually reflects the momentum distribution.
As we have already mentioned, the realistic interaction demands 
HO functions  with large $Q$ in the coordinate space.
Noting that the Fourier transform of the HO function
in the coordinate space is again the HO function in the momentum space,
the HO functions with large $Q$ certainly contain large-momentum components. 
Refs.~\cite{schiavilla07,horiuchi07,suzuki08,feldmeier11}
showed that the momentum distribution has a long tail 
due to the tensor and short-range correlations.
The HO functions with large $Q$ play a role in enhancing
the high momentum tail of the momentum distribution, whereas
those with small $Q$ describe the mean-field structure
below the Fermi momentum.

Since the inclusion of all the high-$Q$ components
is not practical for heavier nuclei,
an effective interaction starting from the realistic $NN$ interaction is usually employed
to accelerate the convergence. Such effective interactions are derived 
in several approaches, for example, Lee-Suzuki transformation~\cite{suzuki80},
unitary correlation operator method 
(UCOM)~\cite{feldmeier98, neff03}, and
similarity renormalization group~\cite{bogner07}.
A softened interaction always improves 
the energy convergence~\cite{roth10,bogner10}
and succeeds to reproduce some low-lying spectra 
of light nuclei. See Ref.~\cite{barrett13} 
for many such applications in the NCSM framework.

\subsection{First excited state of $^4$He: cluster correlation}
\label{excited.sec}

\begin{figure}[ht]
\begin{center}
\epsfig{file=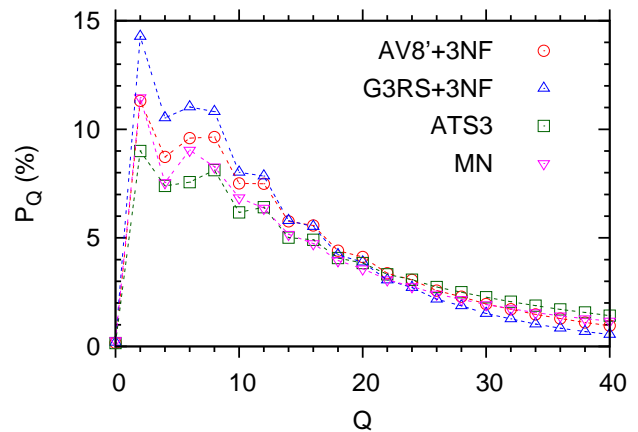,scale=1.35}
\caption{(Color online) 
Occupation probability distributions 
of the total HO quanta for 
the first excited state of $^4$He.}
\label{HOquanta4Heext.fig}
\end{center}
\end{figure}

The $P_Q$ distribution of the excited state of $^4$He  
shows a pattern quite different from that of the ground state.
Figure~\ref{HOquanta4Heext.fig} plots
$P_Q$ of the first excited $J^\pi T=0^+0$ state
calculated with the four interaction models.
The $P_{0}$ value almost vanishes, obviously because 
the state is orthogonal to the ground state whose major configuration 
is $(0s)^4$.  The distribution is 
less interaction-dependent at $Q<10$
compared to that of the ground state, which appears to be attributed to  
the weakly bound $3N+N$ cluster structure of the first 
excited state~\cite{hiyama02,horiuchi08}.
Assuming that the scattering length between $3N$ and $N$ 
is much larger than its effective range, 
the system does not depend much on the detail of the interaction.
This universal property is found in atomic systems
and its similarity to the first excited $0^+0$ state 
is discussed in Ref.~\cite{hiyama12}.
Beyond $Q=10$, $P_Q$ decreases monotonously and very slowly with 
increasing $Q$, and 
the values of $M_Q$ and $\sigma_Q$ in the case of the AV8$^\prime$+3NF potential 
turn out to be 15.3 and 13.3, respectively. 
Appreciable probability still exists 
even at $Q=30$, 
which is too large for standard shell-model calculations 
to incorporate~\cite{jurgenson11}.
From the angular momentum decomposition of $P_Q$ we 
find out that the $L=0$ component, $P^0_Q$, dominates over the whole $Q$ region.
This is also consistent with the fact that the $0_2^+$ state 
has an $S$-wave $3N+N$ cluster structure. If a state has a cluster structure, 
its $P_Q$ distribution spreads over large $Q$ because  
describing the relative motion between the clusters up to the asymptotic region  
requires configurations with large $Q$, even though  
the intrinsic wave functions of the clusters do not contain high HO excitations~\cite{suzuki96}. 
Other well-known examples, which support this fact, include   
the Hoyle state of $^{12}$C~\cite{suzuki96,neff09} and the first excited state of 
$^{16}$O~\cite{suzuki76,horiuchi14}. 

\begin{figure}[ht]
\begin{center}
\epsfig{file=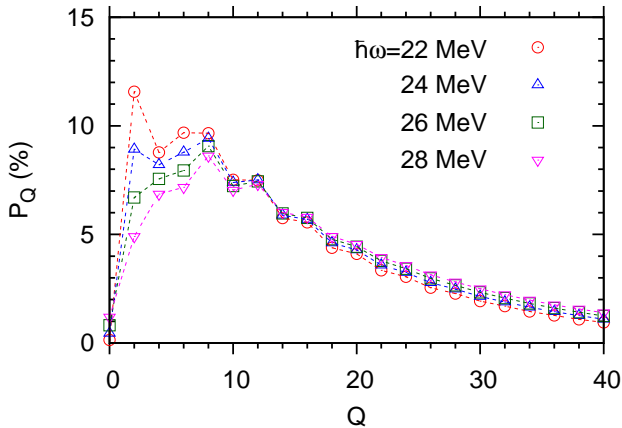,scale=1.35}
\caption{(Color online) $\hbar\omega$ dependence
of the occupation probability of the total HO quanta
for the first excited state of $^4$He. 
The AV8$^\prime$+3NF potential is used.}
\label{HOhwdepAV8p4Heext.fig}
\end{center}
\end{figure}

One may think that the first excited state of $^4$He 
can be described well in a shell model by choosing 
$\hbar\omega$ appropriately. To examine this question more closely, we exhibit 
the $\hbar\omega$ dependence of $P_Q$ in Fig.~\ref{HOhwdepAV8p4Heext.fig}. 
The probability for $Q < 10$ depends on $\hbar\omega$, 
but no practical dependence is found beyond this $Q$ value.
Since 
the occupation probability is still significant for
$Q>8$, we conclude that no appropriate choice for $\hbar\omega$ exists
to describe the cluster state in the conventional shell-model truncation. 
Since the maximum major shell in
shell-model calculations cannot be
taken sufficiently large at present, it 
is reasonable to improve the wrong asymptotic behavior of the HO basis 
by combining with some other methods such as the resonating group 
method~\cite{baroni13,quaglioni13}.

\subsection{Inversion doublets in $^4$He}
\label{inversion.sec}

As shown in Ref.~\cite{horiuchi08}, 
the first excited state of $^{4}$He has those 
negative-parity partners that have 
basically the same intrinsic structure.  
If a system has a two-cluster structure consisting of 
asymmetric subsystems, both 
positive and negative parity states may be found 
around the relevant threshold energy. A well-known example is $^{16}$O 
with a $^{12}$C$+\alpha$ structure~\cite{suzuki76, horiuchi14}.   As a `mini' version of $^{16}$O\,
the spectrum of $^{4}$He has some similarity to that of $^{16}$O.
Because of the spin-isospin coupling of $3N+N$ clusters, 
seven negative-parity states appear in $^4$He above 
the first excited $0^+0$ state, as shown in 
calculations with the AV8$^\prime$+3NF potential~\cite{horiuchi13a}.

\begin{figure}[ht]
\begin{center}
\epsfig{file=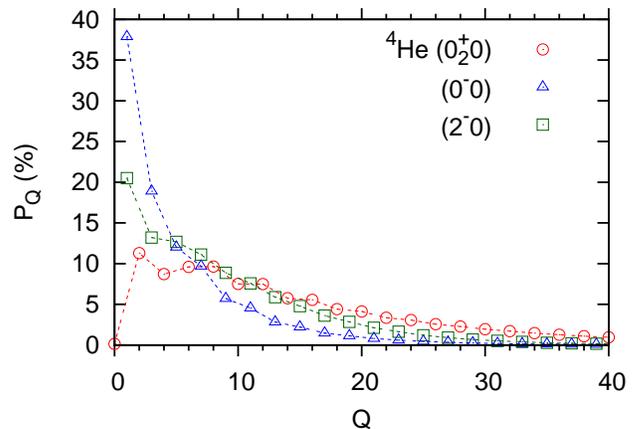,scale=1.35}
\caption{(Color online) Occupation probability distributions of the HO quanta for 
the excited states of $^4$He. The AV8$^\prime$+3NF potential is used.}
\label{4Heinversion.fig}
\end{center}
\end{figure}

Figure~\ref{4Heinversion.fig} plots the $P_Q$ values of 
the $0^-0 \,(21.01 \,\text{MeV})$ and $2^-0 \,(21.84 \,\text{MeV})$
states that are identified as the parity-inverted partners~\cite{horiuchi08}.
For the sake of comparison, the $P_Q$ of the positive-parity partner, 
the $0^+_20$ state, is also drawn. 
Though the HO occupation probability is widely distributed to high $Q$ values, 
the $P_{1}$ value of the $0^-$ state is 37.9\%, as well as the 
$M_Q$ and $\sigma_Q$ values are 5.42 and 6.36, respectively. 
These values are not as large as the corresponding values for the $0^+_20$ state.
Since it has a significant overlap with the 1p-1h configurations, 
the $0^-0$ state is expected to be described fairly well in large-scale  
shell-model calculations. Compared to the $0^-0$ state, 
the $P_Q$ distribution of the $2^-0$ state is closer to 
that of the $0^+0$: the $P_{1}$ value is 20.5\%, while the 
$M_Q$ and $\sigma_Q$ values grow to 8.72 and 8.00, respectively. Because of the 
$P$-wave centrifugal barrier between $3N$ and $N$ clusters, 
the $2^-0$ state shrinks compared to the $0^+_20$ state and consequently 
the distribution of $P_Q$ is shifted to lower $Q$ values than that 
of the $0^+_20$ state.

\section{Summary}
\label{summary.sec}

We have formulated a method for calculating 
the occupation probability of the number of total harmonic-oscillator (HO) 
quanta $Q$ to shed light on various types of nuclear correlations. 
We have analyzed 
the occupation probability distributions of the precise wave functions 
of $s$-shell nuclei that are obtained in the correlated Gaussian basis 
employing four kinds of interactions.

The HO probability distributions show quite different behavior
reflecting the characteristics of the interaction employed.
In the case of the ground state of $^4$He, 
the tensor force significantly enhances the probability below $Q=20$. 
The short-range repulsion also plays an important 
role in mixing configurations with 
more than $Q=10$ excitations.
For the excited states of $^{4}$He,
the occupation probability is widely distributed to large $Q$ values 
and does not depend so much on the detail of the interaction.
Configurations with a higher number of HO quanta are needed to describe the 
tail of the relative motion between the $3N$ and $N$ clusters. 
In conformity to the parity-inverted doublet structure, 
the similarity of the HO distribution 
of the first excited $J^\pi T=0^+0$ state to that of  
the negative-parity excited states with $0^-0$ and $2^-0$ is discussed.

We find that all the probability distributions beyond $Q=10$ 
are insensitive to the choice for the HO oscillator 
frequency $\hbar\omega$.
These high-$Q$ components in the wave function always exist irrespective 
of whether the interaction is effective or realistic and thereby lead to 
the difficulty or extremely slow convergence in describing the cluster 
structure in the HO basis. 
The analysis presented here is useful for confirming that the occupation 
probability distribution 
in fact reflects important correlations and various kinds of structure 
of the nuclear wave functions. This analysis will be useful for 
providing a hint on how to 
develop an improved truncation scheme for huge shell-model spaces. 

\section*{Acknowledgments}

The authors are greatly indebted to K. Launey for her careful 
reading of the manuscript.
This work was supported in part by JSPS KAKENHI Grant No. 
24540261 and No. 25800121.

\appendix

\section{Matrix element for the projection operator of number of HO quanta}

We define the Jacobi coordinate and the corresponding reduced mass as  
\begin{align}
\bm{x}_i=\frac{1}{i+1}\sum_{j=1}^{i}\bm{r}_{j}-\bm{r}_{i+1},
\quad \mu_i=\frac{i}{i+1}m
\end{align}
with $i=1,\dots, N-1$, 
where $\bm{r}_j$ is the $j$th nucleon coordinate.  
Letting $\bm{\pi}_i$ denote the momentum conjugate to $\bm{x}_i$,
the HO Hamiltonian $H_{\rm HO}(i)$ in Eq.~(\ref{HOope.eq}) reads
\begin{align}
H_{\rm HO}(i)=
\frac{\bm{\pi}_i^2}{2\mu_i}+\frac{1}{2}\mu_i\omega^2\bm{x}_i^2.
\end{align}

We evaluate the matrix element of $\text{e}^{i\theta \mathcal{O}}$ 
between the generating functions of the CG~(\ref{gCG.eq}) in three steps. 
First, we rewrite the generating function in a multiple-integral form 
of a product of Gaussian wave-packets. Next, we act with
$\text{e}^{i\theta \mathcal{O}}$ on the Gaussian wave-packets. 
Finally, the multiple-integral is performed analytically, which leads to 
the required matrix element. 

Let $\psi_{\bm{R}_i}^{\gamma_i}(\bm{x}_i)$ denote a Gaussian wave-packet centered
at $\bm{R}_i$ with a width parameter $\gamma_i$
\begin{align}
\psi_{\bm{R}_i}^{\gamma_i}(\bm{x}_i)=\left(\frac{\gamma_i}{\pi}\right)^{3/4}
\exp\left(-\frac{\gamma_i}{2}(\bm{x}_i-\bm{R}_i)^2\right).
\end{align}
The first step is to use the identity~\cite{varga95}
\begin{widetext}
\begin{align}
g(\bm{s}; A,\bm{x})
&=\left[\frac{(\det\Gamma)^3}{(4\pi)^{N-1}(\det(\Gamma-A))^2}
\right]^{\frac{3}{4}}\exp\Big(-\frac{1}{2}\tilde{\bm{s}}(\Gamma-A)^{-1}\bm{s}\Big) \notag \\
&\quad \times \int d\bm{R}\,g(\Gamma(\Gamma-A)^{-1}\bm{s}; A(\Gamma-A)^{-1}\Gamma,\bm{R})
\prod_{i=1}^{N-1}\psi_{\bm{R}_i}^{\gamma_i}(\bm{x}_i),
\label{genewp.eq}
\end{align}
\end{widetext}
where $\bm{R}$ stands for an ($N-1$)-dimensional column vector whose $i$th element is
$\bm{R}_i$ and $d\bm R=d\bm R_1 d\bm R_2 \ldots d\bm R_{N-1}$.  
$\Gamma$ is an $(N-1)\times (N-1)$  diagonal matrix whose element is chosen to be 
\begin{align}
\Gamma_{ij}=\gamma_i\delta_{i,j}=\frac{\mu_i\omega}{\hbar}\delta_{i,j}.
\end{align} 
The second step is to use the identity (see Eq.~(5) of Ref.~\cite{suzuki96}), which makes 
it possible to obtain 
\begin{align}
&\exp\left(i\theta\left[\frac{1}{\hbar\omega}H_{\rm HO}(j)-\frac{3}{2}\right]\right)
\psi_{\bm{R}_j}^{\gamma_j} (\bm{x}_j) \notag \\
&\quad =\exp\left(-\frac{\gamma_j}{4}(1-z^2){\bm{R}_j^2}\right)
\psi_{z\bm{R}_j}^{\gamma_j} (\bm{x}_j),
\end{align}
where $z=\text{e}^{i\theta}$. The operation of $\text{e}^{i\theta \mathcal{O}}$ on the 
product of the Gaussian wave-packets is then given in a simple form:
\begin{widetext}
\begin{align}
\text{e}^{i\theta \mathcal{O}}\prod_{i=1}^{N-1}\psi_{\bm{R}_i}^{\gamma_i}(\bm{x}_i)
=\exp\left(-\frac{1}{4}(1-z^2)\tilde{\bm R}\Gamma \bm R \right) 
\left(\frac{{\rm det}\Gamma}{\pi^{N-1}}\right)^{3/4} \exp\left(-\frac{1}{2}\tilde{\bm x}\Gamma \bm x+z\tilde{\bm R}\Gamma \bm x -
\frac{1}{2}z^2 \tilde{\bm R}\Gamma \bm R \right).
\end{align}
\end{widetext}
The third step for obtaining $e^{i\theta\mathcal{O}}g(\bm{s};A,\bm{x})$ is to substitute the above result into Eq.~(\ref{genewp.eq}) and integrate over 
$\bm R$, which leads to the following compact result expressed again in terms of the generating 
function~(\ref{gCG.eq}) of CG: 
\begin{align}
&\text{e}^{i\theta\mathcal{O}}g(\bm{s};A,\bm{x})\notag \\
&\ =\left(\frac{\det\Gamma}{\det C}\right)^{3/2}\exp\left(-\frac{1}{2}\tilde{\bm{s}}G
\bm{s}\right)g(z\Gamma C^{-1}\bm{s};D,\bm{x}),
\end{align}
where the matrices $C, \, D$, and $G$ are given by 
\begin{align}
C&=A+\frac{1+z^2}{2}(\Gamma-A),\notag\\
D&=\left(A+\frac{1-z^2}{2}(\Gamma-A)\right)C^{-1}\Gamma,\\
G&=-\frac{1-z^2}{2}C^{-1}.\notag
\end{align}
It is easy to derive Eq.~(\ref{formulaQ.eq}) using the result above. 

A calculation of the matrix element of $\text{e}^{i\theta\mathcal{O}}$ between 
the CG~(\ref{GVR.eq}) is therefore reduced to that of the overlap 
matrix element of the CG.  See Refs.~\cite{svm,suzuki08,aoyama12} that
 detail this process. 
An explicit form for the matrix element reads 
\begin{widetext}
\begin{align}
&\left<F_{(L_3L_4)LM}(u_3,u_4,A^\prime,\bm{x})|
{\rm e}^{i\theta \mathcal{O}}
|F_{(L_1L_2)LM}(u_1,u_2,A,\bm{x})\right>\notag\\
&\quad =\left(\frac{{\rm det}\,\Gamma\ {\rm det}\,(A+A^\prime)}
{{\rm det}B\,{\rm det}C}\right)^{3/2} 
\left<F_{(L_3L_4)LM}(u_3,u_4,A^\prime,\bm{x})|F_{(L_1L_2)LM}(u_1,u_2,A,\bm{x})
\right>_{\rho_{ij}\to X_{ij}},
\end{align}
\end{widetext}
where $\left<F_{(L_3L_4)LM}(u_3,u_4,A^\prime,\bm{x})|F_{(L_1L_2)LM}(u_1,u_2,A,\bm{x})\right>$ is 
the overlap matrix element (see Eq.~(B.10) of Ref.~\cite{suzuki08}) and the
$\rho_{ij}=\tilde{u}_i(A+A')^{-1}u_j$, which appears in Ref.~\cite{suzuki08},
should be replaced by $X_{ij}$ defined as follows: 
\begin{align}
X_{12}&=\tilde{u}_1\left\{\frac{1-z^2}{2}
C^{-1}+z^2C^{-1}\Gamma G^{-1}\Gamma C^{-1}\right\}u_2, \notag \\
X_{13}&=z\tilde{u}_1C^{-1}\Gamma G^{-1}u_3,\notag \\
X_{14}&=z\tilde{u}_1C^{-1}\Gamma G^{-1}u_4,\notag \\
X_{23}&=z\tilde{u}_2C^{-1}\Gamma G^{-1}u_3,\notag \\
X_{24}&=z\tilde{u}_2C^{-1}\Gamma G^{-1}u_4,\notag \\
X_{34}&=\tilde{u}_3G^{-1}u_4.
\end{align}

\end{document}